\documentclass[a4paper,11pt]{article}
\usepackage{pos,epsfig,caption,subcaption}
\newcommand{\EE}{{\cal E}}
\def\blue#1{{\color{blue}#1}}
\def\green#1{{\color{green}#1}}
\def\red#1{{\color{red}#1}}

\title{Eight loop form factors, amplitudes and patterns in planar $\mathcal{N}=4$ super-Yang-Mills theory}
\ShortTitle{Eight loop form factors, amplitudes and patterns in planar $\mathcal{N}=4$}

\author*[a]{Lance J.~Dixon}
\author[a]{Zhenjie Li}

\affiliation[a]{SLAC National Accelerator Laboratory,
  Stanford, CA 94309, USA}

\emailAdd{lance@slac.stanford.edu}
\emailAdd{munuxilee@gmail.com}

\abstract{The simplest nontrivial amplitude in planar $\mathcal{N}=4$ super-Yang-Mills
theory is six-gluon scattering in the maximally-helicity-violating configuration.  It has been computed to 8 loops with the help of antipodal duality, which relates it to the three-point form factor of a protected operator, the chiral stress tensor super-multiplet, represented also as ${\rm tr} \phi^2$.  In this talk, we describe the computation to 8 loops of another three-point form factor, for the operator ${\rm tr}\phi^3$.   This form factor lives in the same restricted
space of polylogarithms as the ${\rm tr}\phi^2$ form factor.   We also report on all-order patterns for sequences of coefficients in the symbols of these
polylogarithmic results, for the leading discontinuity of the ${\rm tr}\phi^3$ form factor.}

\FullConference{Loops and Legs in Quantum Field Theory (LL2026)\\
12-17, April, 2026\\
Bayreuth, Germany\\}

\begin{document}
\maketitle

\section{Introduction}
\label{sec:intro}

The planar (large $N_c$) limit of $\mathcal{N}=4$ super-Yang-Mills theory (SYM) is the prototype conformal field theory in the Anti-de-Sitter/conformal field theory (AdS/CFT) duality.
It has played a key role in developing modern scattering amplitude methods. The large $N_c$
limits leads to quantum integrability, which gives rise to the quantum spectral curve, the duality of amplitudes and Wilson loops, and the pentagon operator product expansion (OPE) for Wilson loops~\cite{Basso:2013vsa}.
There is also a mysterious principle of maximal transcendentality relating $\mathcal{N}=4$ SYM to quantum chromodynamics (QCD)~\cite{Kotikov:2001sc,Kotikov:2002ab,Kotikov:2004er}, which was first seen in the DGLAP evolution of twist-two operators, as well as in BFKL (small $x$) evolution.

Beyond twist-two and BFKL anomalous dimensions, planar $\mathcal{N}=4$ SYM teaches us ``something'' about the leading transcendental part of QCD amplitudes, although the connection is not always precise.  For simple processes, the $\mathcal{N}=4$ answer is a polylogarithmic function of known transcendental weight (weight $2L$ at loop order $L$) and high loop order computations can be relatively simple.  Furthermore,
we know how to remove all infrared divergences, either via ``framing'' dual Wilson loops~\cite{Basso:2013vsa}, or via a BDS(-like) ansatz~\cite{Bern:2005iz,Alday:2009dv,Dixon:2015iva,Caron-Huot:2016owq}.  

In this talk we will consider the three-point form factors for the BPS operators called ${\rm tr}\phi^2$ and ${\rm tr}\phi^3$.  The former is also a representative of the chiral stress tensor multiplet.  The dimensionally-regulated matrix elements of these operators with three massless partons can be divided by a suitable BDS-like ansatz~\cite{Dixon:2020bbt,Dixon:2022rse,Basso:2024hlx}, which removes all infrared divergences, leaving the finite functions $\EE_3(u,v)$ for ${\rm tr}\phi^2$ and $\EE_{3,3}(u,v)$ for ${\rm tr}\phi^3$.  Here $u$ and $v$ are the underlying dimensionless kinematic variables, expressed in terms of the massless parton momenta $p_i$, $i=1,2,3$, with $p_i^2=0$:
\begin{equation}  
u = \frac{(p_1+p_2)^2}{(p_1+p_2+p_3)^2}\,, \qquad 
v = \frac{(p_2+p_3)^2}{(p_1+p_2+p_3)^2}\,, \qquad 
w = \frac{(p_3+p_1)^2}{(p_1+p_2+p_3)^2} = 1 - u - v\,.
\label{eq:uvwdef}
\end{equation}
The operator momentum is $p_1+p_2+p_3$.

The loop expansion of these functions is
\begin{equation}
 \EE_{3[,3]}(u,v) = \sum_{L=0}^\infty g^{2L} \,\EE_{3[,3]}^{(L)}(u,v),
\end{equation}
where $g^2 \equiv N_c g_{YM}^2/(4\pi)^2$. (We have also normalized by the tree-level form factor, so that $\EE_{3[,3]}^{(0)}\equiv1$.)  These quantities are ``Goldilocks'' processes for the planar $\mathcal{N}=4$ SYM form-factor/amplitude bootstrap because they are simple, but not too simple.  They depend on only two dimensionless variables via weight $2L$ polylogarithms that involve only six ``letters'', a remarkably small symbol alphabet~\cite{Goncharov:2010jf}. The simplicity of the polylogarithmic function space makes it possible to construct an ansatz for the answer to high loop order.  The ansatz can then be subjected to a variety of constraints, including the generalization of the pentagon OPE to form factors (FFOPE)~\cite{Sever:2020jjx,Sever:2021nsq,Sever:2021xga,Basso:2023bwv} which controls the limit $v\to0$ at fixed $u$.   This combination of factors made it possible to get to 8 loops for ${\rm tr} \phi^2$~\cite{Dixon:2022rse}. With the discovery of antipodal duality~\cite{Dixon:2021tdw} between this form factor and the maximally-helicity violating (MHV) six-gluon amplitude~\cite{Caron-Huot:2019vjl}, it was then possible to use the form factor to compute the MHV amplitude to 8 loops as well~\cite{Dixon:2023kop}.

In this contribution, we will review some of these developments and also present new results for the three-point ${\rm tr} \phi^3$ form factor to 8 loops~\cite{DixonLiToAppear}; previously it was known to 6 loops~\cite{Basso:2024hlx}. It was also computed at 2 loops analytically~\cite{Brandhuber:2014ica}, and at 3 loops numerically~\cite{Lin:2024pki,Lin:2021qol} and analytically~\cite{Henn:2024pki}.
Thus the total Loops\,+\,Legs for these processes is currently 8$\,+\,4 = 12$ (for the two form factors, counting the operator momentum) and 8$\,+\,6 = 14$ (for the MHV six-gluon amplitude). 

It is noteworthy that another representative of the chiral stress tensor multiplet is ${\rm tr} \, G_{\mu\nu,{\rm SD}} G^{\mu\nu}_{\rm SD}$, where SD stands for the self-dual part of the field strength tensor.  This operator is very similar to the operator ${\rm tr} \, G_{\mu\nu} G^{\mu\nu}$ that couples the Higgs boson to gluons, as the leading term in the limit of a large top quark mass.  It has been known for a while that through two loops, the leading transcendental part of the QCD result coincides with the $\mathcal{N}=4$ result~\cite{Brandhuber:2012vm,Duhr:2012fh}.
Recently, the leading-color QCD result was computed at three loops~\cite{Chen:2025utl}; comparing with the $\mathcal{N}=4$ result~\cite{Dixon:2020bbt}, it was found that the principle of maximal transcendentality still holds for this process at three loops and large $N_c$.

\section{Polylogarithmic function space and ansatz}
\label{sec:fnspace}

We bootstrap the three-point form factors of both ${\rm tr}\phi^2$ and ${\rm tr}\phi^3$ using the same function space ${\cal C}$.  (The fact that the same space works for both form factors is not obvious.)
We map weight $n$ polylogarithms $F$ to their symbols~\cite{Goncharov:2010jf} by iterating their derivatives,
\begin{align}
dF &= \sum_{s_i\in {\cal L}} F^{s_i} \, d \ln s_i
\quad\Rightarrow\quad {\cal S}[F] = \sum_{s_i\in {\cal L}} {\cal S} [ F^{s_i} ] \otimes s_i \,,
\label{symboldef1}\\
dF^{s_j} &= \sum_{s_i\in {\cal L}} F^{s_i,s_j} \, d \ln s_i
\quad\Rightarrow\quad {\cal S}[F] = \sum_{i,j} {\cal S} [ F^{s_i,s_j} ] \otimes s_i \otimes s_j \,,
\label{symboldef2}
\end{align}
reaching finally
\begin{equation}
{\cal S}[F] = \sum_{s_{i_1},s_{i_2},\ldots,s_{i_n}}  F^{i_1,i_2,\ldots,i_n} \, s_{i_1} \otimes s_{i_2} \otimes \cdots \otimes s_{i_n}  \,,
\label{symboldef}
\end{equation}
where the symbol letters $s_i$ belong to the symbol alphabet ${\cal L}$.  The tensor $F^{s_{i_1},s_{i_2},\ldots,s_{i_n}}$ is a set of rational numbers --- in fact, it turns out to always be integers for $\EE_3^{(L)}$ and $\EE_{3,3}^{(L)}$ through 8 loops!

The symbol alphabet for the ${\rm tr}\phi^2$ and ${\rm tr}\phi^3$ three-point form factors is
\begin{equation}
{\cal L} = \{ a,b,c,d,e,f \}  \,,
\label{Ldef}
\end{equation}
where 
\begin{align}
    &a = \sqrt{\frac{u}{vw}} \,, \quad 
    b = \sqrt{\frac{v}{wu}} \,, \quad
    c = \sqrt{\frac{w}{uv}} \,, \nonumber\\
    &d = \frac{1-u}{u} \,, \quad
    e = \frac{1-v}{v} \,, \quad
    f = \frac{1-w}{w} \,. \quad
\end{align}
The quantities $\EE_3^{(L)}$ and $\EE_{3,3}^{(L)}$ are invariant under a dihedral symmetry $D_3 \equiv S_3$ that comes from permuting the three massless momenta $p_i$.  Its generators act on the alphabet by
\begin{align}
    \label{eq:cycle}
   {\bf cycle:}&\ \ a \to b \to c \to a, \qquad d \to e \to f \to d, \\
    \label{eq:flip}
    {\bf flip:}&\ \ a \leftrightarrow b, \qquad d \leftrightarrow e.
\end{align}

The symbols of the ${\rm tr}\phi^2$ form factor at one and two loops are just one and two terms, plus their dihedral images:
\begin{align}
 {\cal S}[ \EE_3^{(1)} ] &= -2 \ b \otimes d\ +\  \text{dihedral images},  \label{SEE3_1}\\
 {\cal S}[ \EE_3^{(2)} ] &= 8\ b \otimes d \otimes d  \otimes d\ + \ 16\ b \otimes b \otimes b  \otimes d\ +\ \text{dihedral images}.  \label{SEE3_2}
\end{align}
The symbols of the ${\rm tr}\phi^3$ form factor at one and two loops are not quite as simple:
\begin{align}
 {\cal S}[ \EE_{3,3}^{(1)} ] &= a\otimes b - a\otimes a\ +\  \text{dihedral images},  \label{SEE33_1}\\
 {\cal S}[ \EE_{3,3}^{(2)} ] &= 
6\ {\tt aaaa}-6\ {\tt aaba}-6\ {\tt abaa}+5\ {\tt abba}+{\tt abca}-3\ {\tt aeaa}+3\ {\tt aeca}+8\ {\tt aabb}-6\ {\tt aaab} \nonumber\\
&\hskip0.4cm
-6\ {\tt abbb}+5\ {\tt abab}
+{\tt abcb}+{\tt acab}+{\tt accb}-2\ {\tt aacb}-2\ {\tt acbb}+3\ {\tt afab}-3\ {\tt afbb}\nonumber\\
&\hskip0.4cm
 \ +\ \text{dihedral images},  \label{SEE33_2}
\end{align}
where we abbreviated
$a\otimes b\otimes c \otimes a \rightarrow {\tt abca}$, etc., in eq.~\eqref{SEE33_2} to save space.

The amplitude/form factor bootstrap has been refined over the years~\cite{Dixon:2011pw,Dixon:2014voa,Drummond:2014ffa,Dixon:2020bbt,Dixon:2022rse,He:2025tyv}.
For the ${\rm tr}\phi^3$ form factor, it combines:
\begin{enumerate}
\item Knowledge of the polylogarithmic function space (borrowed from experience with ${\rm tr}\phi^2$).
\item Empirical constraints on multiple final entries (analogous to ${\rm tr}\phi^2$ but different in detail).
\item Knowledge of leading and next-to-leading discontinuities from the FFOPE~\cite{Basso:2023bwv}, but lifted into general kinematics.
\item Generation of an ansatz as a linear combination of basis elements in the function space, consistent with the above constraints, including dihedral invariance.
\item Matching the near-collinear limit to the FFOPE prediction for one flux-tube excitation, but all subleading logarithms~\cite{Basso:2023bwv}, in order to fix all remaining constants in the ansatz.
\end{enumerate}

We now expand on these five elements of the bootstrap.

\noindent
{\bf 1. Function Space:} Besides the symbol alphabet~\eqref{Ldef}, there are branch-cut constraints: The only physical discontinuities are at $(p_i+p_j)^2=0$ and at $(p_1+p_2+p_3)^2=0$.  They imply that every word in the symbol starts with $a$, $b$ or $c$, because $d=0$ corresponds to $(p_1+p_2)^2 = (p_1+p_2+p_3)^2$, where there is no such discontinuity.  The same branch-cut constraint has implications deeper in the symbol, namely that~\cite{Dixon:2020bbt} 
\begin{equation}
  F^d |_{v,w\to0} = 0.
\label{more_branch_cut}
\end{equation}
There are also powerful constraints on pairs of adjacent letters. They include integrability, which enforces the equality of mixed partial derivatives, $d^2F = 0$. Other constraints ban the appearance of the following pairs of adjacent letters:
\begin{equation}
\ldots {\tt ad} \ldots, \qquad 
\ldots {\tt da} \ldots, \qquad 
\ldots {\tt de} \ldots, \qquad \text{[$+$ dihedral]}
\end{equation}
There is also a constraint on adjacent triples,
\begin{equation}
 F^{a,a,b} + F^{a,b,b} + F^{a,c,b} = 0. 
 \quad \text{[$+$ dihedral]}
\end{equation}
The constraints beyond integrability can be understood from the extended Steinmann relations~\cite{Caron-Huot:2019bsq} or cluster adjacency~\cite{Drummond:2017ssj}, properties which are obeyed by the hexagon function space ${\cal H}$ describing six-gluon amplitudes, after mapping ${\cal H}$ to ${\cal C}$ using antipodal duality~\cite{Dixon:2021tdw}. 
On the other hand, antipodal duality itself is not well understood!

\renewcommand{\arraystretch}{1.25}
\begin{table}[!t]
\centering
\begin{tabular}[t]{l c c c c c c c c c c c c c c c c c}
\hline\hline
weight $n$
& 0 & 1 & 2 & 3 & 4 &  5 &  6 &  7 &  8 & 9 & 10 & 11 & 12
& 13 & 14 & 15 & 16\\\hline\hline
$L=1$
& \green{1} & \blue{2} & \blue{1} &  &  &  &  &  &  &  & & & &  & & &
\\\hline
$L=2$
& \green{1} & \green{3} & 3 & \blue{2} & \blue{1} & & & & & & & & &  & & &
\\\hline
$L=3$
& \green{1} & \green{3} & \green{9} & \blue{13} & \blue{6} & \blue{2} & \blue{1}
&  &  &  & & & &  & & &
\\\hline
$L=4$
& \green{1} & \green{3} & \green{9} & \green{21} & \blue{29} & \blue{13}
& \blue{6} & \blue{2} & \blue{1} &  & & & &  & & &
\\\hline
$L=5$
& \green{1} & \green{3} & \green{9} & \green{21} &  \green{48} & \blue{57}
& \blue{29} & \blue{13} & \blue{6} & \blue{2} & \blue{1} & & &  & & &
\\\hline
$L=6$
& \green{1} & \green{3} & \green{9} & \green{21} & \green{48} & 105
& \blue{112} & \blue{57} & \blue{29} & \blue{13} & \blue{6} & \blue{2}
& \blue{1} &  & & &
\\\hline
$L=7$
& \green{1} & \green{3} & \green{9} & \green{21} & \green{48} & \green{108}
& \green{242} & \blue{206} & \blue{112} & \blue{57} & \blue{29}
& \blue{13} & \blue{6} & \blue{2} & \blue{1} & &
\\\hline
$L=8$
& \green{1} & \green{3} & \green{9} & \green{21} & \green{48} & \green{108}
& \green{242} & 519 & \red{375} & \blue{206} & \blue{112}
& \blue{57} & \blue{29} & \blue{13} & \blue{6} & \blue{2} & \blue{1}
\\\hline\hline
\end{tabular}
\caption{The number of independent $\{n,1,1,\ldots,1\}$ coproducts
  of the form factor $\EE_{3,3}^{(L)}$ through $L=8$ loops, at
  \emph{symbol} level.  The meaning of the colors is discussed in the text.}
\label{tab:EEcopdimsymb}
\end{table}

\noindent
{\bf 2. Multiple final entries:}
After computing the weight $2L$ function $\EE_{3,3}^{(L)}$, we take its iterated derivative, or rather its $\{2L-1,1\}$ coproducts~\eqref{symboldef1}, then its $\{2L-2,1,1\}$ coproducts~\eqref{symboldef2}, and so on. We record the dimension of these spaces in table~\ref{tab:EEcopdimsymb}.
We observe that they {\it saturate} in two different ways: 
\begin{itemize}
    \item We display a number in \green{green} once it appears at the same weight $n$ and at two consecutive loop orders.  This indicates the stabilization of the {\it frontspace} ${\cal C}$.  We also check that the independent functions (or symbols) we find are the same as those obtained by taking coproducts of the ${\rm tr}\phi^2$ form factor $\EE^{(L)}$.
    \item We display a number in \blue{blue} once it appears at weight  $n = 2L-w_f$ for the same $w_f$ at two consecutive loop orders.  Here we can read off the dimensions of the ${\rm tr}\phi^3$ {\it backspace}, and correspondingly, a set of {\it multi-final-entry} relations that enforce these dimensions.
\end{itemize}

For example, we find that there are 4 relations among the 6 possible single final entries of $\EE_{3,3}$:
\begin{equation}
\EE_{3,3}^d = \EE_{3,3}^e = \EE_{3,3}^f = 0, \qquad
\EE_{3,3}^a + \EE_{3,3}^b + \EE_{3,3}^c = 0,
\label{EE33finalentryrelations}
\end{equation}
leaving only the \blue{2} entries recorded in table~\ref{tab:EEcopdimsymb}, say $\EE_{3,3}^a$ and $\EE_{3,3}^b$.
These relations should be contrasted with the 3 relations among the single final entries of the ${\rm tr}\phi^2$ form factor $\EE_3$:  $\EE_3^a = \EE_3^b = \EE_3^c = 0$.
The 6 double final entry relations for $\EE_{3,3}$ are:
\begin{align}
\EE_{3,3}^{d,a} &= \EE_{3,3}^{e,b} = 0, \qquad
\EE_{3,3}^{f,b} = - \EE_{3,3}^{f,a}, \qquad
\EE_{3,3}^{c,a} = - \EE_{3,3}^{a,a} - \EE_{3,3}^{b,a},
\label{EE33doublefinalentryrelations1}\\
\EE_{3,3}^{c,b} &= - \EE_{3,3}^{a,b} - \EE_{3,3}^{b,b}, \qquad 
\EE_{3,3}^{d,b} = \EE_{3,3}^{e,a} - \EE_{3,3}^{f,a} + \EE_{3,3}^{a,b} - \EE_{3,3}^{b,a}.
\label{EE33doublefinalentryrelations2}
\end{align}
They relate $6\times 2$ possible pairs of double final entries (after taking into account the single final entry conditions), and so they leave
\blue{6} independent ones in table~\ref{tab:EEcopdimsymb}.  After each new loop is computed, we learn a new layer of multi-final-entry conditions (at one larger value of $w_f$).  For example, at $L=3$ we see the \blue{6} corresponding to eqs.~\eqref{EE33doublefinalentryrelations1} and \eqref{EE33doublefinalentryrelations2} for the first time; at $L=4$ the \blue{6} is confirmed.

\noindent
{\bf 3. Leading Discontinuity:}
Near the collinear limit where $p_1\parallel p_3$, $w\to0$, we parametrize the kinematics by
\begin{equation}
u = \frac{1}{1+S^2+T^2}\ ,
\qquad v = \frac{S^2}{\left(1+T^2\right)\left(1+S^2+T^2\right)}\ ,
\qquad w = \frac{T^2}{1+T^2}\ ,
\label{OPEparamST}
\end{equation}
with $S = e^\sigma$, $T = e^{-\tau}$.
At $L$ loops, the leading discontinuity of the three-point form factor for ${\rm tr}\phi^3$ (or more precisely the dual periodic Wilson loop $\mathcal{W}_{3,3}$) follows from the FFOPE~\cite{Basso:2023bwv,Basso:2024hlx},
in the limit $T \to 0$:
\begin{equation}
\mathcal{W}^{(L)}_{3,3}|_{T^1 \, \ln^L T} 
= \int_{-\infty}^\infty du \frac{\pi}{\cos(\pi i u)} S^{2 i u}
  \frac{1}{L!} [ 2 (\psi(\tfrac{1}{2}+iu)+\psi(\tfrac{1}{2}-iu)-2\psi(1)) ]^L \,,
\label{LLintu}
\end{equation}
where in this equation $u$ is the rapidity of the single flux-tube excitation, and $\psi(x) = d\ln\Gamma(x)/dx$.

To evaluate~\eqref{LLintu}, we close the $u$ contour in the upper half-plane, sum over residues at $iu = \tfrac{1}{2}+n$, for $n$ a non-negative integer,
and then reconstruct the result as $S/(1+S^2)$ multiplied by a linear combination of weight $L$ harmonic polylogarithms (HPLs)~\cite{Remiddi:1999ew}
with indices 0 and 1 and argument $S^2$.
Normally such an expression would only hold near $T=0$, but by a suitable choice of parametrization we find that the leading- (and next-to-leading-) discontinuities can be resummed to all powers in $T$ so that they hold in the bulk. Their symbol involves two combinations of the usual 6 letters:
\begin{equation}
    {\tt A} = \frac{u}{v} = \frac{1+T^2}{S^2}\,, \qquad {\tt C} = \frac{1-w}{v} = \frac{1+S^2+T^2}{S^2} \,.
\label{ACdef}    
\end{equation}

After converting the leading discontinuity from the normalization $\mathcal{W}_{3,3}$
to the normalization ${\cal R} \equiv \exp(R_{3,3})|_{T^1\,\ln^L T}$, where $R_{3,3}$ is the remainder function,
we find for the symbols of ${\cal R}^{(L)}$:
\begin{align}
{\cal R}^{(0)} &= 1, \\
{\cal R}^{(1)} &= 0, \\
{\cal R}^{(2)} &= - \, {\tt AA}, \\
{\cal R}^{(3)} &= -2 \, {\tt CAA} - 6 \, {\tt ACA} + 4 \, {\tt AAA}, \\
{\cal R}^{(4)} &=
-16\,{\tt CCAA}-16\,{\tt CACA}-48\,{\tt ACCA}
+16\,{\tt CAAA}+32\,{\tt ACAA}+32\,{\tt AACA}
-15\,{\tt AAAA}, \\
{\cal R}^{(5)} &= 
-160\,{\tt CCCAA}-160\,{\tt CACCA}-160\,{\tt CCACA}
-480\,{\tt ACCCA}+160\,{\tt CCAAA}+160\,{\tt CACAA} \nonumber\\
&\hskip0.5cm
+160\,{\tt CAACA}+320\,{\tt ACCAA}+320\,{\tt ACACA}
+320\,{\tt AACCA}-102\,{\tt CAAAA}-150\,{\tt ACAAA} \nonumber\\
&\hskip0.5cm
-190\,{\tt AACAA}-150\,{\tt AAACA}+56\,{\tt AAAAA}\,.
\end{align}
We computed up to ${\cal R}^{(10)}$ from \eqref{LLintu}. Next we observed~\cite{DixonLiToAppear} the following recursive pattern:
Let $c_L({\tt XCA}^k)$ be the coefficient in ${\cal R}^{(L)}$ of the string {\tt XCA}$^k$, where {\tt X} is any string of {\tt A}'s and {\tt C}'s of length $L-k-1$. In other words,
for any given string at $L$ loops, we find the last {\tt C} in the string,
which always has $k\geq1$ {\tt A}'s behind it. (That is, $c_L({\tt X})=0$ if {\tt X} ends in a {\tt C}.) Then,
\begin{equation}
c_L({\tt XCA}^k) = 2 L \, c_{L-1}({\tt XA}^k)
  - \xi_k \, \frac{L!}{(L-k-1)!} \, c_{L-k-1}({\tt X}),
\label{mainrecurs}
\end{equation}
where the coefficients $\xi_k$ can be expressed in terms of a confluent hypergeometric function,
\begin{eqnarray}
\xi_k &=& \sum_{m=0}^k \frac{(-1)^{k-m} \, 2^{m+1}}{(m+1)!} \frac{k!}{m!(k-m)!} \\
      &=& 2 \,(-1)^k \ {}_1F_1(-k,2;2) \\
      &=& 0, -\tfrac{2}{3}, \tfrac{2}{3}, -\tfrac{2}{5}, \tfrac{4}{45},
      \tfrac{10}{63}, -\tfrac{32}{105}, \tfrac{142}{405}, \ldots \,.
\label{xik}
\end{eqnarray}
If we normalize the coefficients by $L!$, so that
\begin{equation}
\hat{c}_L({\tt XCA}^k) = \frac{c_L({\tt XCA}^k)}{L!} \,,
\label{chatdef}
\end{equation}
then they are no longer always integers,
but the recursion relation becomes simpler:
\begin{equation}
{\hat c}_L({\tt XCA}^k) = 2 \, {\hat c}_{L-1}({\tt XA}^k)
  - \xi_k  \, {\hat c}_{L-k-1}({\tt X}) \,.
\label{chatmainrecurs}
\end{equation}
Because the recursion always lowers the number of {\tt C}'s in a string
by 1, we also record the boundary condition for the recursion, which
is the string with no {\tt C}'s:
\begin{equation}
\hat{c}_L({\tt A}^L) = \frac{c_L({\tt A}^L)}{L!}
= \sum_{m=0}^L \frac{(-1)^{L-m}}{m!} \frac{L!}{m!(L-m)!}
= (-1)^L \, {}_1F_1(-L,1;1) \,,
\end{equation}
so ${}_1F_1$ appears a second time.  Note that the boundary condition also obeys a recursion relation,
\begin{equation}
\hat{c}_L({\tt A}^L) = - \frac{L-1}{L}
\Bigl[ 2 \hat{c}_{L-1}({\tt A}^{L-1}) + \hat{c}_{L-2}({\tt A}^{L-2}) \Bigr] \,.
\end{equation}
We have verified that the next-to-leading discontinuity (the coefficient of $T^1 \, \ln^{L-1}T$) can also be lifted into the bulk, where it also depends on {\tt A} and {\tt C}, and on a logarithmic letter.

\noindent
{\bf 4. Ansatz:}
We build an ansatz using a previous construction of the ${\cal C}$ frontspace through weight 8~\cite{Dixon:2022rse}, where it has dimension 1,495 including zeta-valued terms (i.e.~terms beyond the symbol).
At eight loops, these 1,495 functions are married to a 403-dimensional weight 8 backspace for ${\rm tr}\phi^3$ via a {\it sewing matrix}~\cite{Basso:2024hlx}. (Only after the fact do we know that there are only \red{375} independent weight 8 backspace functions.)
The sewing-matrix ansatz has 1,495$\,\times \, 403 =$ 602,485 parameters to be determined by the constraints.
They were solved by the {\sc SparseRREF} package~\cite{LiPrimeSolver}. Only one 10 digit prime was needed to reconstruct the entire answer over the rational numbers.  The rational reconstruction was greatly aided by the fact that the 1,251$\,\times\,403 =$ 504,153 symbol-level coefficients were all integers (in the right representation).

\noindent
{\bf 5. Matching to FFOPE:} We matched the ansatz to the predictions of the FFOPE through 8 loops, at the level of a single flux tube excitation but all subleading logarithms~\cite{Basso:2023bwv,Basso:2024hlx}. It is straightforward to evaluate the flux-tube formula in a fashion similar to \eqref{LLintu}, as a truncated sum over residues, which provides a truncated series expansion in $S$.  However, the reconstruction in terms of HPLs at the full weight 16 (with over $2^{16}$ terms) would require too many terms in the series expansion, if not for patterns that we found in the lower-weight results that relate the coefficients of many different HPLs~\cite{DixonLiToAppear}.

\section{Results}
\label{sec:results}

We fixed the coefficients uniquely in the ansatz for the ${\rm tr}\phi^3$ form factor through 8 loops, first at symbol level and then at function level.  We then integrated up the result along various lines through the phase space~\cite{Dixon:2022rse,Basso:2024hlx}, which requires fixing boundary conditions involving multiple zeta values at weights from 9 up to 16.
Weight 16 multiple polylogarithms take too long to evaluate, so we performed series expansions along lines with overlapping radii of convergence.  

\begin{figure}[t]
\centering
 \begin{subfigure}[b]{0.45\textwidth}
         \centering
         \rotatebox{90}{\tiny \hspace{0.12\paperwidth}\clap{$\frac{\EE_{3,3}^{(L)}(u,u,1-2u)}{\EE_{3,3}^{(L-1)}(u,u,1-2u)}$}}
         \includegraphics[height=0.23\paperwidth,trim=6.5cm 0.24cm 3.3cm 0.28cm,clip]{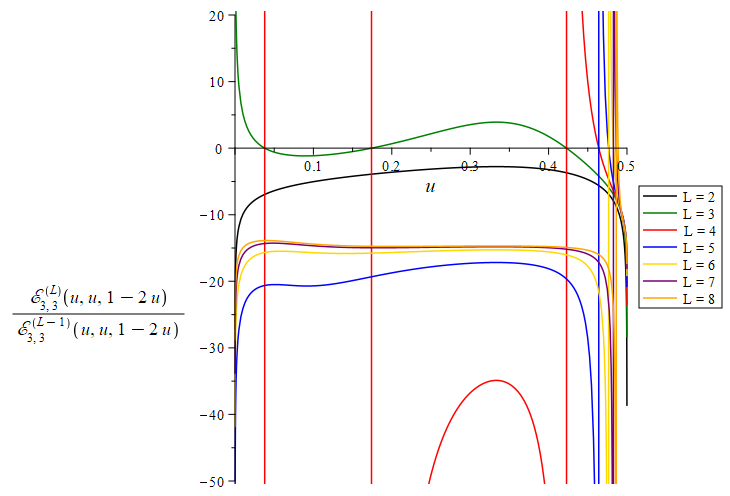}
         \caption{\phantom{.}}
         \label{Fig:EE33_sym8}
     \end{subfigure}
     \hfill
      \begin{subfigure}[b]{0.54\textwidth}
         \centering
         \rotatebox{90}{\tiny \hspace{0.12\paperwidth}\clap{$\bigg|\frac{\EE_{3,3}^{(L)}(r)}{\EE_{3,3}^{(L-1)}(r)}\bigg|$}}
         \includegraphics[height=0.23\paperwidth,trim=4.1cm 0.2cm 0.1cm 0.34cm,clip]{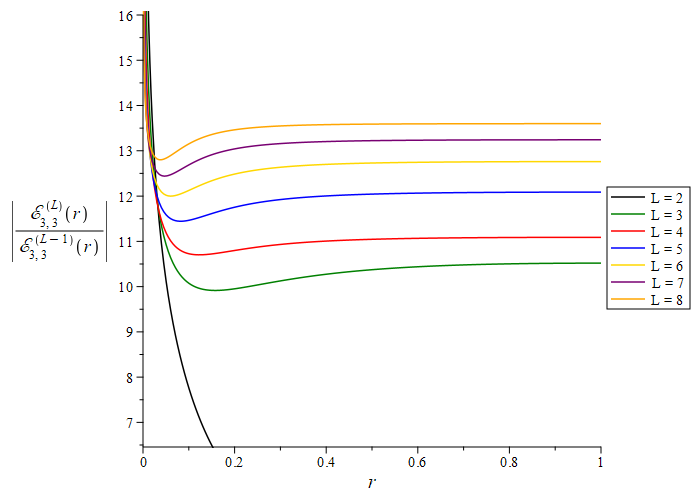}
         \caption{\phantom{.}}
         \label{Fig:EE33_fa_r8}
     \end{subfigure}
\caption{(a) Ratios of successive loop orders for the finite function $\EE_{3,3}^{(L)}$ along the $v=u$ line within the Euclidean region, which corresponds to $0 < u < 1/2$.  (b) Ratios of $\EE_{3,3}^{(L)}$ at successive loop orders in the high-energy fixed-angle region, $u,v\to \infty$ with $r=u/v$ fixed, within the space-like scattering region.}
\label{Fig:EE338loops}
\end{figure}

In figure~\ref{Fig:EE33_sym8}, we plot the ratio of $\EE_{3,3}^{(L)}$ to the previous loop order
along the symmetric line $v=u$ (where $(u,v,w)=(u,u,1-2u)$) in the (pseudo-)Euclidean (decay) region $0 < u < \tfrac{1}{2}$.  For $L=6,7,8$, the ratio is very close to constant, away from the endpoints where $\EE_{3,3}^{(L)}$ becomes singular.
The motivation for considering successive loop ratios is the expected finite radius of convergence of the perturbative series for planar ${\cal N}=4$ SYM.  Because its beta function vanishes, this theory lacks renormalons.  Because of the large $N_c$ limit, instantons are suppressed.  These ``-ons'' are the leading reasons why a perturbative series in quantum field theory is asymptotic, not convergent.  Furthermore, the flux-tube representation of the amplitude or form factor is built off the same vacuum that is responsible for the integrable representation of the cusp anomalous dimension~\cite{Beisert:2006ez}, so we might expect the same radius of convergence as for $\Gamma_{\rm cusp}$, namely $1/16$ for coupling $g^2 \equiv N_c\, g_{\rm YM}^2/(4\pi)^2$.
This radius of convergence corresponds to a large-order growth,
\begin{equation}
\frac{\Gamma_{\rm cusp}^{(L)}}{\Gamma_{\rm cusp}^{(L-1)}} \to -16, \qquad \text{as $L\to\infty$,}
\label{cuspasympt}
\end{equation}
and $(-16)$ is approximately the plateau value seen in figure~\ref{Fig:EE33_sym8}.

In figure~\ref{Fig:EE33_fa_r8} we plot the same ratio of successive form factors $\EE_{3,3}^{(L)}$, but now in the high-energy, fixed-angle limit where the operator momentum-squared is negligible, namely $u,v\to\infty$ with the ratio $r = u/v$ held fixed.  Again we see that the ratio of successive loop orders is quite independent of $r$, away from $r=0$ which is a forward scattering limit.   On the other hand, the successive-loop ratio in this region is still changing significantly with $L$, even at $L=8$.  

\begin{figure}[t]
\centering
 \begin{subfigure}[b]{0.45\textwidth}
         \centering
         \rotatebox{90}{\tiny \hspace{0.12\paperwidth}\clap{$\frac{\EE^{(L)}(u,u,1)}{\EE^{(L-1)}(u,u,1)}$}}        
         \includegraphics[height=0.23\paperwidth,trim=5.7cm 0.24cm 2.3cm 0.28cm,clip]{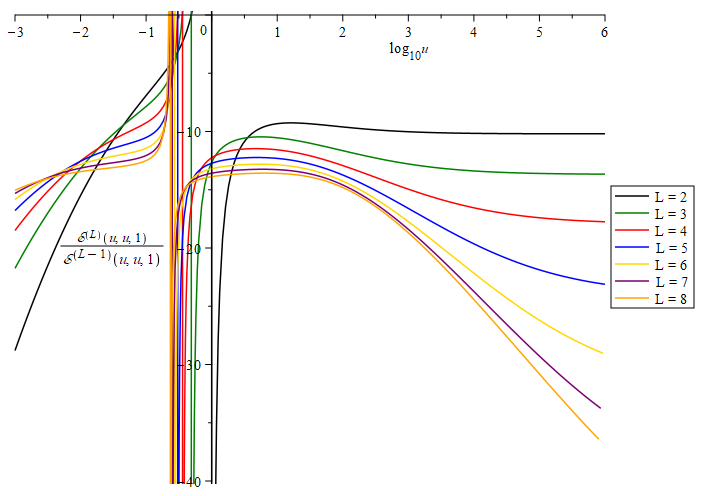}
         \caption{\phantom{.}}
         \label{Fig:EZMHVuu1_8}
     \end{subfigure}
     \hfill
      \begin{subfigure}[b]{0.54\textwidth}
         \centering
         \rotatebox{90}{\tiny \hspace{0.12\paperwidth}\clap{$\frac{\EE^{(L)}(u,1,1)}{\EE^{(L-1)}(u,1,1)}$}}
         \includegraphics[height=0.23\paperwidth,trim=4.72cm 0.2cm 0.1cm 0.34cm,clip]{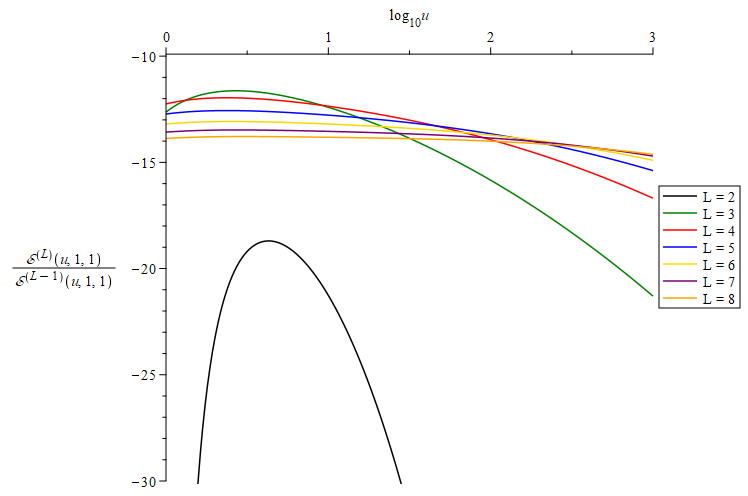}
         \caption{\phantom{.}}
         \label{Fig:EZMHVu11_8}
     \end{subfigure}
\caption{(a) Ratios of successive loop orders for the finite function $\EE^{(L)}(u,v,w)$ associated with the MHV 6-gluon amplitude along the line $(u,v,w)=(u,u,1)$ within the Euclidean region, through $L=8$.  (b) The same ratios on the line $(u,1,1)$. From ref.~\cite{Dixon:2023kop}.}
\label{Fig:EZMHV_8}
\end{figure}

Figures~\ref{Fig:EZMHVuu1_8} and \ref{Fig:EZMHVu11_8} illustrate that there is a similar lack of stabilization~\cite{Dixon:2023kop} of ratios for the MHV 6-gluon amplitude $\EE^{(L)}(u,v,w)$ at large values of the cross-ratio $u$ along two different lines. 
The cross ratios are double ratios of Mandelstam invariants,
\begin{equation}
 u = \frac{s_{12}s_{45}}{s_{123}s_{345}} \,, \qquad 
v = \frac{s_{23}s_{56}}{s_{234}s_{456}} \,, \qquad w = \frac{s_{34}s_{61}}{s_{345}s_{561}} \,.
\label{crossratio6}
\end{equation}
%

The computation of $\EE^{(L)}$ through seven loops was performed using a conventional bootstrap~\cite{Caron-Huot:2019vjl}. At eight loops, the computation was assisted~\cite{Dixon:2023kop} by antipodal duality~\cite{Dixon:2021tdw}, which predicts the symbol on the two-dimensional {\it parity-preserving} surface
\begin{equation}
\Delta(u,v,w) \equiv (1-u-v-w)^2-4uvw = 0,
\label{Deltaeq0}
\end{equation}
given the symbol of the ${\rm tr}\phi^2$ three-point form factor~\cite{Dixon:2022rse}.
Matching to the value on $\Delta=0$ is a very powerful constraint, which will be indispensable in pushing the MHV 6-gluon amplitude beyond eight loops, i.e.~to reach Loops\,+\,Legs $= 9 + 6 = 15$.

\vskip0.5cm 

\noindent
{\bf Acknowledgments}

We thank Benjamin Basso for several useful conversations.
This work was supported in part by the U.S. Department of Energy (DOE) under Awards No.~DE-FOA-0002705 [KA/OR55/22 (AIHEP)] and~DE-AC02-76SF00515.


\end{document}